\documentclass{article}
\usepackage{PRIMEarxiv}

\usepackage{amsmath,amssymb,amsfonts}
\usepackage{algorithmic}
\usepackage{graphicx, xcolor, comment, enumerate, multirow, multicol}
\usepackage{textcomp}
\usepackage[square,sort,comma,numbers]{natbib}
\usepackage{xcolor}
\usepackage{xspace}
\usepackage{hyperref}
\usepackage{multicol}
\usepackage{multirow}
\usepackage{booktabs}

\newcommand{\X}{\mathcal{X}\xspace}
\newcommand{\Y}{\mathcal{Y}}
\newcommand{\x}{\mathbf{x}}
\newcommand{\y}{\mathbf{y}}

\usepackage{url}

\begin{document}

\fancyhead[LO]{Combining Euclidean Alignment and Data Augmentation for BCI decoding}
  
%% Title 
\title{Combining Euclidean Alignment and Data Augmentation for BCI decoding}

\author{Gustavo H. Rodrigues$^{1}$ \quad Bruno Aristimunha$^{2, 3}$ \quad Sylvain Chevallier$^{2}$ \quad Raphael Y. de Camargo$^{1}$  \\
\\
\small $^1$ University of São Paulo, Sao Paulo, Brazil, \\
\small $^2$ Université Paris-Saclay, Inria TAU team, LISN-CNRS, France, \\
\small $^3$ Federal University of ABC (UFABC), Santo André, Brazil,\\
}

\maketitle

\begin{abstract}
Automated classification of electroencephalogram (EEG) signals is complex due to their high dimensionality, non-stationarity, low signal-to-noise ratio, and variability between subjects. Deep neural networks (DNNs) have shown promising results for EEG classification, but the above challenges hinder their performance. Euclidean Alignment (EA) and Data Augmentation (DA) are two promising techniques for improving DNN training by permitting the use of data from multiple subjects, increasing the data, and regularizing the available data. In this paper, we perform a detailed evaluation of the combined use of EA and DA with DNNs for EEG decoding. We trained individual models and shared models with data from multiple subjects and showed that combining EA and DA generates synergies that improve the accuracy of most models and datasets. Also, the shared models combined with fine-tuning benefited the most, with an overall increase of 8.41\% in classification accuracy.
%Automated classification of electroencephalogram (EEG) signals is complex due to their high dimensionality, non-stationarity, low signal-to-noise ratio, and variability between subjects. Although deep neural networks have shown promising results for automated classification, training these models still faces challenges. These include the small amount of data to train models and the data irregularities between subjects. In this paper, we evaluate the combination of Euclidean Alignment with data augmentation techniques to improve the performance of deep neural network architectures for EEG decoding. We trained individual models and shared models with multiple subjects and showed that the addition of EA and data augmentation contributed to improving the performance of both model types, particularly in the latter case. 
\end{abstract}


\section{Introduction}

%How does this support your intro message? 

The brain-computer interface (BCI) is an emerging technology of recent decades, which establishes a direct link between external devices and brain signals \cite{BCI}. %BCI is emerging
Traditionally, BCIs have been used for medical applications such as emotion recognition, human-machine interactions, and, above all, neural control of prosthetic limbs \cite{rehabilitation_Jin2022, man_machine_Liu2021, emotions_Li2022}. %Application of BCI
Among the various non-invasive techniques, the electroencephalogram (EEG) is widely used to detect neural activity since it is portable and has a low acquisition cost. Despite these advantages, decoding EEG is still challenging due to the inherent complexity of this type of data, resulting from the high dimensionality and low signal-to-noise ratio of this type of signal. %Challanges of BCI

To overcome the challenges of EEG decoding, researchers started to evaluate the use of Deep Neural Networks (DNNs), which can extract important features directly from raw data and have been shown to provide superior performance compared to conventional methods~\cite{DOSE2018532, eegnet, Roy2019, Schirrmeister2017, paillard2024green}. Yet, despite the promising results of DNNs, they require large training datasets. Two possibilities for increasing the availability of data are training DNNs using data from multiple subjects and using data augmentation techniques. 

Data augmentation~\cite{NIPS2012_c399862d, augmentation_Lotte2015} increases the training sets' size by adding new synthetic examples by performing transformations in the data while preserving the labels. This strategy alleviates the problem of overfitting by providing more training data while also providing regularization since the DNNs become invariant to the imposed transformation. It has shown that data augmentation improves the accuracy of EEG Decoding~\cite{rommel2022benchmark, Ali2022}.

The second way to improve the training of DNNs is to use data from multiple subjects. Considering each subject as a domain, we must achieve greater similarity between domains~\cite{Pan2010}. One approach is to use Euclidean alignment (EA)~\cite{He2020:euclidean}, which is a pre-processing method that centers the average of the covariance matrices of each subject EEG trials on the identity matrix. Although little explored in the context of neural networks, this type of alignment has already shown good results in recent studies \cite{He2020:euclidean} and is generating increasing interest due to its computational efficiency and simplicity.

Although these techniques have been studied separately~\cite{augmentation_Lotte2015, song2022eeg, rommel2022benchmark, rommel2022cadda}, using both techniques together can generate positive interactions. For instance, the regularization provided by the augmentation process may enhance the domain adaptation provided by Euclidean alignment.

In this paper, we evaluate the combined use of Euclidean alignment and a data augmentation strategy, called Segmentation and Reconstruction (S\&R), for training DNNs using data from single or multiple subjects. Our results show that combining S\&R with EA improved the accuracy of most models and datasets, especially in the case of shared models, reaching up to $13\%$ of improvement with fine-tuning. 

%We organize the paper as follows. First, section 2 introduces the relevant related work in transfer learning, EEG decoding and cognitive tasks structure investigation. In section 3, we formalize transfer learning for EEG decoding. Then, section 4 outlines our experimental setting, including a description of the two EEG datasets and their respective decoding modalities, as well as our training and data processing protocols. Subsequently, we present the main results of our experiments in section 5, highlighting the significant improvements in decoding performance achieved through transfer learning. Finally, in section 6, we discuss the broader implications of our findings and offer insights into the hierarchical and asymmetric relations between cognitive tasks.
       
\section{Material and Methods}

\subsection{EEG Decoding}

We assume EEG decoding is a machine-learning classification problem when evaluating our proposal \cite{king:2020}. %, $f: \X \rightarrow \Y$. 
We consider signals as real-valued matrices $\X$, where each matrix $\x_{i}\in\X$ is associated with a cognitive task $\y_i \in \Y$. 
The matrix $\x_{i}$ is an element of the set $\X = \mathbb{R}^{C \times T}$, with $C$ as the number of channels (electrodes), $T$ is the number of time points in the matrices and $i \in \{ 1\ldots N\}$ are in the index vector, with $N$ as the total number of trials/matrices. 

The decoding problem involves learning a neural network model to map each matrix $\x$ to the associated label $\y$, in other words, $f_\theta: \X \to \Y$. 
The model parameters $\theta$ are optimized to minimize the average loss function $\ell$ across the training dataset:

\begin{equation}\label{eq:learning-pb}
	\min_\theta \frac{1}{N} \sum \ell( f_\theta (\x_i), \y_i) \enspace .
\end{equation}

In all our experiments, the loss function $\ell$ used is the balanced cross-entropy. 
To analyze the robustness of the data augmentation with the Euclidean Alignment, we employed four well-established neural networks for EEG decoding, the \emph{ShallowNet}~\cite{Schirrmeister2017},  \emph{Deep4Net}~\cite{Schirrmeister2017}, \emph{EEGNet}~\cite{eegnet} and \emph{EEGConformer}~\cite{song2022eeg} implemented on \textsc{braindecode} library v0.7.

\subsection{Datasets}

We selected three motor imagery datasets to validate our approach, BCI Competition IV 2a \cite{bnci2014001}, the Cho2015 \cite{Cho2017}, and Shin2017A \cite{Shin2017}, available at the \textsc{MOABB} v1.0 library \cite{aristimunha_2023}. 
We chose these large datasets, in the number of trials or the number of subjects, to avoid the common BCI practice of small and single dataset bias~\cite{jayaram2018moabb, chevallier2024largest}. 

The first dataset, denoted in the rest of the paper as \textit{BNCI2014001}, consists of 9 subjects, with two sessions (different days) and four cognitive imagery tasks: left hand, right hand, and both feet and tongue movements. 
The second dataset, \textit{Cho2017}, contains 52 subjects with two cognitive imagery tasks,  left-hand and right-hand, recorded with a single session. 
Finally, the \textit{Shin2017A} includes data from 29 subjects registered in three sessions, with two cognitive imagery tasks, left hand and right hand. 

Regarding the preprocessing steps applied to the datasets, we conduct a re-scaling voltage from V to $\mu$V, band-pass filtering with interval $[4-38]$ Hz, exponential moving standardization \cite{engemann2022reusable}, artifact rejection based on annotations, and baseline correction when necessary. These steps were designed and aligned with recent work on the importance of pre-processing to separate the marks from the central nervous and peripheral systems \cite{Bomatter2023}.

\subsection{Euclidean Alignment}

The natural variability of brain signals represents a significant obstacle in developing a model that can be effectively generalized to several individuals or even the same subject between different days. 
To address this challenge, we employ the Euclidean alignment (\emph{EA}) technique \cite{he2019transfer, junqueira2024systematic}. For each subject's run with $N$ trials, we calculate the arithmetic mean $\bar{R}$ of the covariance matrices:

\begin{equation}
	\bar{R} = \frac{1}{N}\sum_{i=1}^{N} \x_i \x_i^{\top},
\end{equation}

\noindent Next, we perform the alignment as the square root of the inverse of $\bar{R}$ and apply it back to each matrix.

\begin{equation}
	\tilde{X}_i = \bar{R}^{-1/2}\x_i,
\end{equation}

\noindent with these steps, the average covariance matrices in the data become equal to the identity matrix. 
Consequently, the distributions of the covariance matrices of different subjects become more similar, a desirable aspect of improving the model’s generability.

\subsection{Segment \& Reconstruction -- S\&R}

While the alignment technique can be seen as a domain generation component, the data augmentation is considered a regularization transformation. In EEG decoding, data augmentation has been known to enhance the brain decoding task ~\cite{rommel2022cadda, rommel2022benchmark, augmentation_Lotte2015}. 

Because of the non-stationarity property present in the signal, the most important segments of the signal do not have a fixed size or point in the temporal dimension. Here, we employ one of the possible data augmentations that capture this component and introduce more variability to the $f$ neural network. 

\begin{figure}[htbp]
	\centerline{\includegraphics[width=0.9\textwidth]{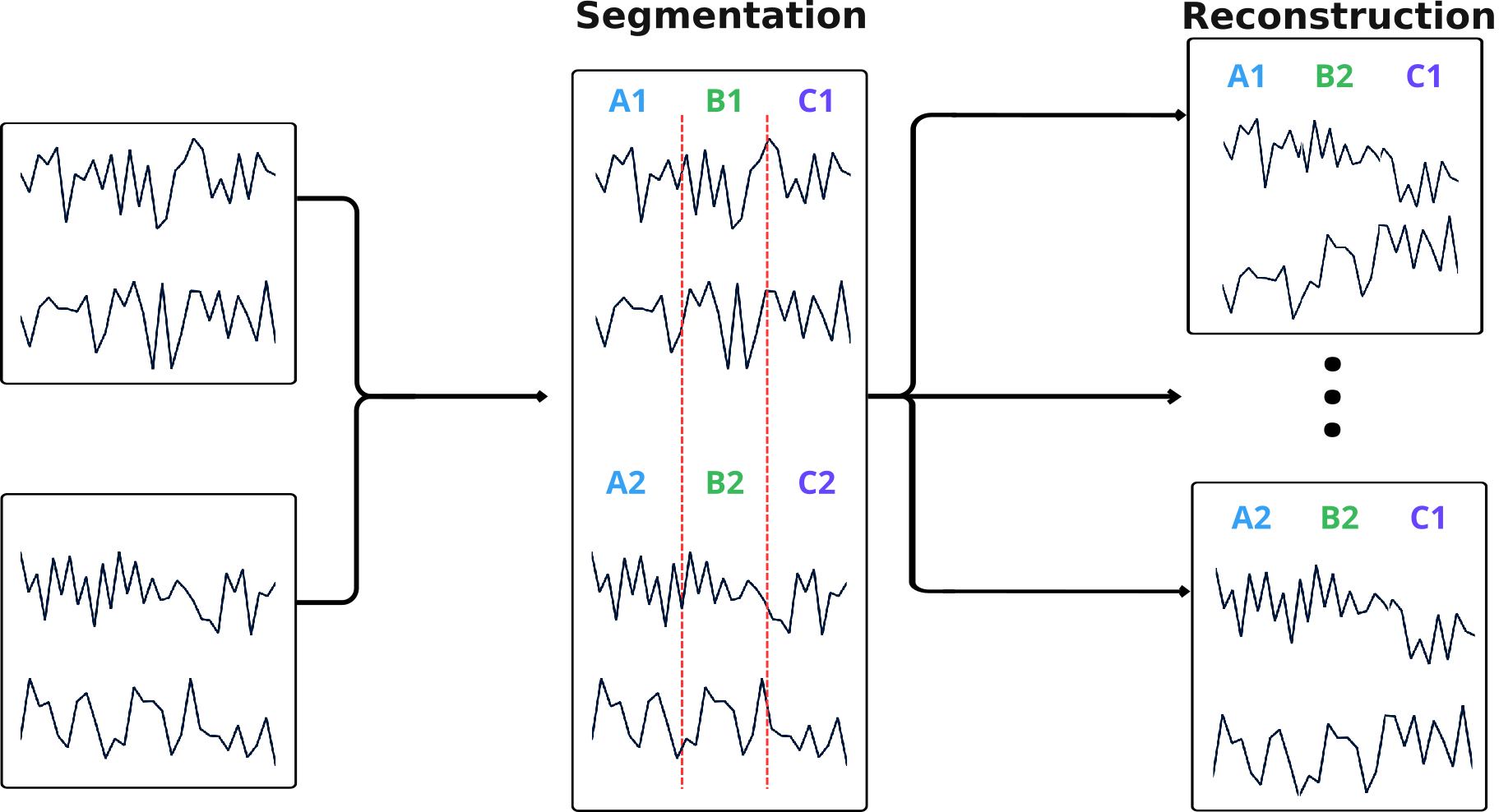}}
	\caption{Principle of synthetic EEG data generation using \textit{Segmentation and Reconstruction} - $S\&R$ strategy in the time domain. The $s$ segmentations are done randomly and the reconstruction preserves the labels of each trial (i.e. there is only a mixing of trials from the same class).}
	\label{augmentation}
\end{figure}

Given the associated $\x$ signal with dimensions $C\times T$ for each class, we \emph{segment} the temporal component into $s$ parts. Then, across the set of trials for each class, we randomly concatenate those segments to form new synthetic \emph{reconstructed} trials, maintaining the original temporal order. An illustration of the process is present in ~\autoref{augmentation}. The Segment and Reconstruction data augmentation increased the number of samples during the batch size of each iteration, with the concatenation of the original trials and the reconstructed synthetic trials.

%However, conventional strategies of adding Gaussian Noise or Random cropping can further decrease the signal-to-noise ratio or destroy the original coherence of the signal. 

\section{Experimental Setup}

Various tests were carried out to compare the accuracies and check whether there were any significant changes when varying the training and data processing conditions, as shown in the graphs in the next subsection. With these results, the performance of the architectures was also compared.

As a first step, the comparative tests conducted initially considered the effects of Euclidean alignment on the models, checking whether or not there was an improvement in the accuracy values for the classes (tasks). In addition, the contribution of data augmentation (S\&R) to the model's performance in the classification task was also checked. After these cases, the model was tested with EA and S\&R together\footnote{Euclidean Alignment processing is realized in an offline fashion, whereas S\&R data augmentation is executed online.}.

\begin{figure}[htbp]
	\centerline{\includegraphics[width=\textwidth]{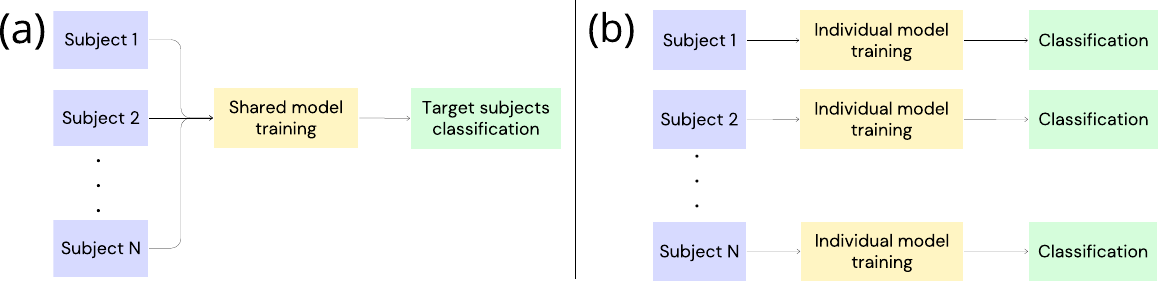}}
	\caption{\textbf{(a)} Cross-subject training, $5$-folds cross-validation over the subjects data. \textbf{(b)} Individual training. Data from a single subject is used for training and validation.}
	\label{individual-all}
\end{figure}

To examine its performance, we analyzed the models in two main situations. Firstly, we used individual models, in which we evaluated the architecture performance per subject (\autoref{individual-all}\textbf{b}). In this case, for \textit{BNCI2014\_001}, we used one session for training and another for testing.
For \textit{Cho2017}, which has 200 trials per subject, we used 70\% of data for train, 15\% for validation, and 15\% for testing. 
Finally, for \textit{Shin2017A}, we used the first session for train, and the other two for validation and test, respectively.

Secondly, the model was trained with data from all the subjects. The strategy was based on choosing a random batch of subjects as the targets, which will be the test set, and the other subjects will be used for training (\autoref{individual-all}\textbf{a}). For the experiments with fine-tuning, we followed the same split as in the individual models with the test subjects sets for \textit{BNCI2014001} and \textit{Shin2017A}. In this case, the train data became the fine-tuning set, and the remaining was used as the test set. For \textit{Cho2017}, we split the target subject set in half, using the first part for fine-tuning and the other for testing. The results obtained follow in the next section.

For both modalities (individual and shared), the models have trained $200$ epochs, using an early stopping of 75 iterations, learning rate of $0.000625$, weight decay of $0.1$, and a batch size of $64$ trials\footnote{In the case of individual training with \textit{Shin2017A} dataset, due to the limited number of trials, we used a batch size of 20.}. The loss criterion and optimizer used were CrossEntropyLoss and Adam, respectively, with beta values equal to $0.5$ and $0.999$. For fine-tuning, the models were trained for an additional 100 epochs, with a patience of 30. In all cases, we used $s$ in S\&R data augmentation set to 12. We used a one-tailed paired permutation test with and without data transformation, and we combines the p-values resulting from each dataset taken separately via Stouffer \cite{jayaram2018moabb}.

\section{Results And Discussion}

We evaluated the effectiveness of Euclidean Alignment and S\&R data augmentation using individual and shared models. Our aim was to determine whether the use of these transformations, both individually and combined, would help to improve BCI decoding accuracy. 

For the individual models, trained using data from single subjects, using EA, S\&R, or EA and S\&R did not improve the accuracy, as shown in the column overall in ~\autoref{tab1}. When looking for individual models, we note that for EEG-Conformer and ShallowNet, using EA and S\&R improved average performance in $2.53$\% and $4.05$\%. However, for DeepNet, there was a decrease of $8.34\%$ when using EA and S\&R. One possible explanation is that DeepNet is a larger model, which may be overfitting the augmented data. Still, the results indicate that using EA and S\&R for training with data from single subjects does not improve accuracy consistently.

\begin{figure}[htbp]
    \centerline{\includegraphics[width=\textwidth]{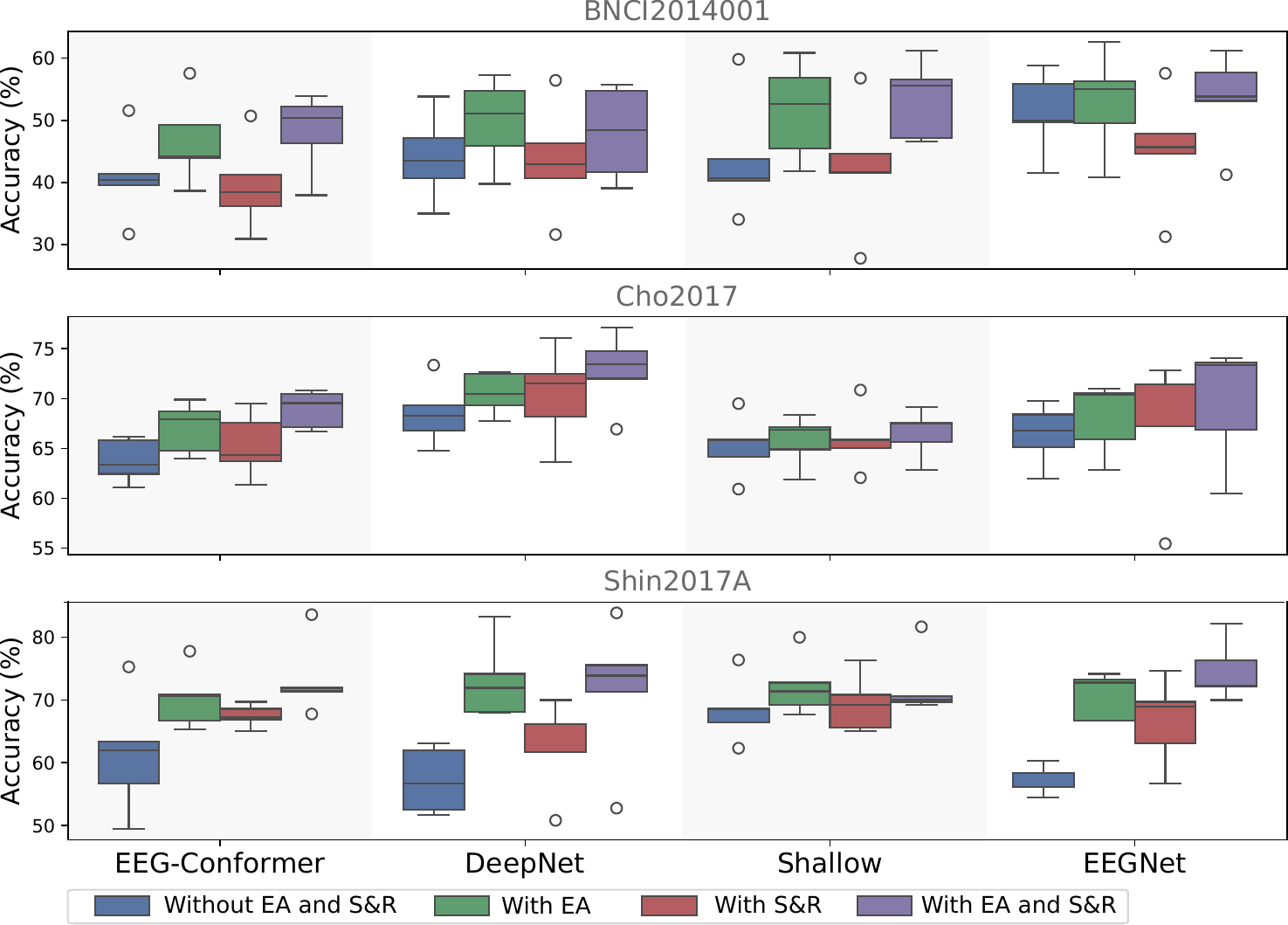}}
    \caption{Prediction accuracy using EEG-Conformer, ShallowNet, DeepNet, and EEGNet shared models with and without Euclidean alignment and data augmentation (S\&R). The statistic comparison between EA+S\&R and without EA and S\&R shows significant difference with $p < 0.001$.}
    \label{cross-subject}
\end{figure}

When training shared models with data from all subjects in a dataset, using either EA, S\&R, and EA and S\&R improved the accuracy of all models compared to the models without EA and S\&R (\autoref{tab1}). Overall, the impact of using EA was the most prominent, but combining EA and S\&R resulted in further improvements. When evaluating individual datasets, shown in \autoref{cross-subject}, using EA and S\&R generated improvements in all cases, but using only S\&R reduced the accuracy in some models in the \textit{BNCI2014001} dataset. 
We can see that using EA and S\&R enabled the shared models to perform similarly to the individual models (\autoref{tab1}), even without using any target subject data for fine-tuning (although we used a small amount of target data to adjust the Euclidean Alignment). Also, the larger model, DeepNet, benefited the most from using data from multiple subjects, enabled by the use of EA and S\&R.

\begin{figure}[htbp]
    \centerline{\includegraphics[width=\textwidth]{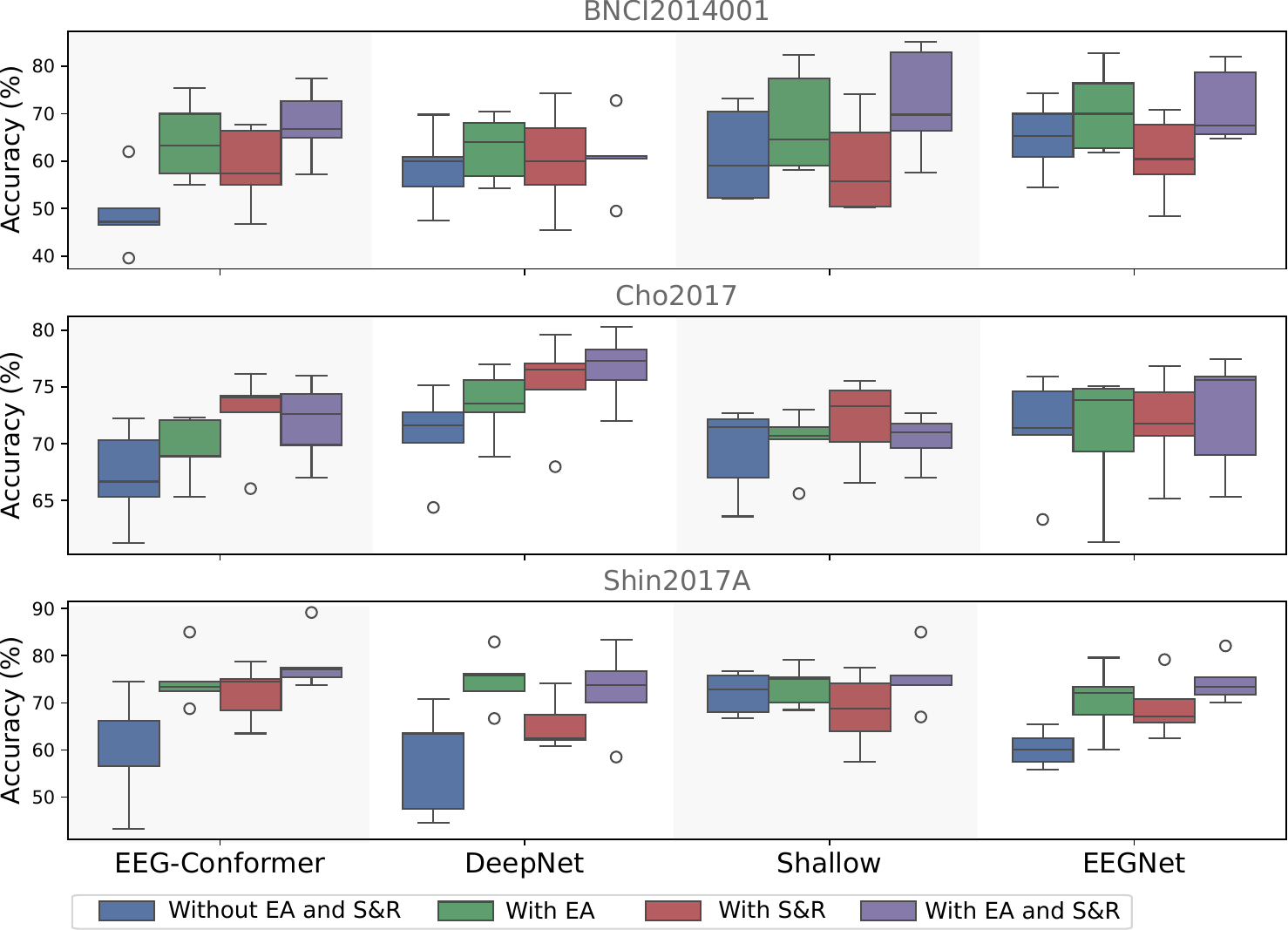}}
    \caption{Prediction accuracy using EEG-Conformer, ShallowNet, DeepNet, and EEGNet shared models after fine-tuning, with and without Euclidean alignment and data augmentation (S\&R). The statistic comparison between EA+S\&R and without EA and S\&R shows significant difference with $p < 0.001$.}
    \label{cross-subject-ft}
\end{figure}

When we fine-tuned the models using target subject data, using EA and S\&R resulted in the most prominent improvements compared to the base model (\autoref{tab1} and \autoref{cross-subject-ft}). Still, using only EA or S\&R individually also improved the accuracies, except for the S\&R with the ShallowNet. Moreover, compared to the individual models, the fine-tuned models with EA, S\&R, or EA and S\&R improved the overall accuracies for all models (\autoref{tab2}). This shows that combining these techniques, EA and S\&R, enabled the effective training of shared models with data from multiple subjects. Finally, combining EA and S\&R generated synergies in the shared and fine-tuned models, resulting in the best overall accuracies, except for DeepNet, which worked better using only EA.

%We obtained improvements up to 9.37\%, 5.92\%, 8.56\% and 7.05\% in average accuracy with EEG-Conformer, DeepNet, ShallowNet and EEGNet, respectively.

\begin{table*}[htbp]
\centering
\caption{Average accuracy contribution of EA and S\&R (mean $\pm$ standard deviation) compared to the baseline model of each training paradigm (Individual, Shared, and Fine-tuned)}
\begin{center}
%\scriptsize
\resizebox{\linewidth}{!}{

\begin{tabular}{l|l|cccc|c}
\toprule
    \multirow{2}{*}{\bf Processing} & \multirow{2}{*}{\bf Modality} &  \multicolumn{4}{c|}{\bf Models} & \multirow{2}{*}{\bf Overall}\\
    \cmidrule{3-6}
& & EEG-Conformer& DeepNet& ShallowNet & EEGNet \\
    \midrule
\multirow{3}{*}{\bf EA} &Individual & 3.37 $\pm$ 9.86 & -3.54 $\pm$ 9.81 & 2.43 $\pm$ 6.83 & -1.06 $\pm$ 9.25 & 0.30 $\pm$ 8.94 \\
&Shared & 5.99 $\pm$ 3.85 & 7.90 $\pm$ 3.64 & 4.20 $\pm$ 2.53 & 5.53 $\pm$ 2.76 & 5.91 $\pm$ 3.19\\
&Shared and Fine-tuning & 10.53 $\pm$ 6.35 & \textbf{7.94} $\pm$ \textbf{5.00} & 3.10 $\pm$ 1.92 & 5.20 $\pm$ 3.64 & 6.69 $\pm$ 4.23\\
\midrule

\multirow{3}{*}{\bf S\&R} &Individual & 0.97 $\pm$ 5.51 & -3.89 $\pm$ 8.46 & 2.10 $\pm$ 6.43 & 1.24 $\pm$ 5.17 & 0.10 $\pm$ 6.39 \\
&Shared & 2.07 $\pm$ 4.47 & 2.40 $\pm$ 4.21 & 0.28 $\pm$ 2.03 & 1.23 $\pm$ 4.71  & 1.50 $\pm$ 3.86\\
&Shared and Fine-tuning & 8.58 $\pm$ 5.31 & 4.55 $\pm$ 3.49 & -1.04 $\pm$ 3.69 & 1.79 $\pm$ 3.01 & 3.47 $\pm$ 3.87 \\
\midrule

\multirow{3}{*}{\bf EA and S\&R} &Individual & 2.53 $\pm$ 7.56 & -8.34 $\pm$ 10.56 & 4.05 $\pm$ 7.78 & -0.28 $\pm$ 8.32 & -0.51 $\pm$ 8.55 \\
&Shared & 8.08 $\pm$ 3.76 & 7.53 $\pm$ 4.27 & \textbf{5.07} $\pm$ \textbf{2.61} & \textbf{7.53} $\pm$ \textbf{3.72} & 7.05 $\pm$ 3.59\\
&Shared and Fine-tuning & \textbf{13.60} $\pm$ \textbf{6.38} & 7.58 $\pm$ 6.58 & 5.00 $\pm$ 3.47 & 7.47 $\pm$ 2.94  & \textbf{8.41} $\pm$ \textbf{4.84}\\
\bottomrule
\multicolumn{7}{l}{\textbf{Bold} for the best model accuracy improvement.}
\end{tabular}}
\label{tab1}
\end{center}
\end{table*}

\begin{table*}[htbp]
\caption{Average accuracy of Individual and Cross-subject with fine-tuning paradigms (mean $\pm$ standard deviation)}
\begin{center}
%\scriptsize
\resizebox{\linewidth}{!}{
\begin{tabular}{c|ccc|ccc}
\toprule
\multirow{2}{*}{\bf Model} & \multicolumn{3}{c|}{\bf Individual} &  \multicolumn{3}{c}{\bf Shared and Fine-tuned} \\
\cmidrule{2-7} 

& \textbf{EA}& \textbf{S\&R}& \textbf{EA and S\&R} & \textbf{EA} & \textbf{S\&R} & \textbf{EA and S\&R}\\
\midrule

EEG-Conformer & 46.53 $\pm$ 12.38 & 48.38 $\pm$ 14.58 & 47.45 $\pm$ 12.53 & 64.2 $\pm$ 7.59 & 58.65 $\pm$ 7.72 & \textbf{67.81} $\pm$ \textbf{6.86}\\

%\hline

DeepNet & 33.68 $\pm$ 8.20 & 32.52 $\pm$ 5.31 & 28.24 $\pm$ 5.90 & \textbf{62.74} $\pm$ \textbf{6.27} & 60.35 $\pm$ 9.88 & 60.90 $\pm$ 7.36\\
%\hline

ShallowNet & 57.87 $\pm$ 10.91 & 54.05 $\pm$ 13.79 & 62.27 $\pm$ 10.29 & 68.30 $\pm$ 9.81 & 59.27 $\pm$ 9.39 & \textbf{72.36} $\pm$ \textbf{10.34}\\
%\hline
EEGNet & 48.26 $\pm$ 10.53 & 49.77 $\pm$ 9.92 & 49.31 $\pm$ 11.39 & 70.70 $\pm$ 7.99 & 60.94 $\pm$ 7.92 & \textbf{71.70} $\pm$ \textbf{7.15}\\
\bottomrule
\multicolumn{7}{l}{\textbf{Bold} for the best model accuracy.}
\end{tabular}}
\label{tab2}
\end{center}
\end{table*}

%Furthermore, these results allow us to consider the technique of cross-subject data useful for classifying EEG signals between different individuals. However, we have also to consider the specificity of data for each subject, which can be seen in the difference of values for individual and shared models, in order to optimize learning and obtain more reliable values, which can make it difficult to generalize models for EEG signals. 

%Furthermore, these results allow us to consider the technique of cross-subject data useful for classifying EEG signals between different individuals. However, we have also to consider the specificity of data for each subject, which can be seen in the difference of values for individual and shared models, in order to optimize learning and obtain more reliable values, which can make it difficult to generalize models for EEG signals. 

Other projects have also explored data augmentation to improve deep learning models in the context of decoding EEG data. For instance, Rommel et al.~\cite{rommel2022benchmark} made a systematic comparison of data augmentation methods in EEG data to improve model performance. They obtained improvements of up to $25\%$ and $45\%$, in low data regimes, for \textit{BNCI2014001} using the \textit{SmoothTimeMask} and \textit{FTSurrogate} strategies, respectively. Ali et al. \cite{Ali2022} used a new data augmentation strategy and its proposed feature extraction method, reaching significant improvements compared to state-of-the-art networks. Our combination of EA and S\&R showed an improvement in the average accuracy of $13.60\%$ for EEG-Conformer, $7.58\%$ for DeepNet, $5.00\%$ for ShallowNet and $7.47\%$ for EEGNet shared models regarding the three datasets.

Similarly, for Euclidean alignment, He and Dongrui \cite{He2020:euclidean} showed an increase of up to $10\%$ when using EA with machine learning algorithms. More recently, Junqueira, B. et al. \cite{junqueira2024systematic} also showed an improvement of $4.33\%$ when applying EA as a preprocessing step for deep learning architectures. We systematically evaluated using both Euclidean Alignment and Data Augmentation in Deep Learning models. We showed that using EA and S\&R enabled the training of shared models with data from all subjects in a dataset, resulting in better overall accuracies in the evaluated datasets. 

For individual models, we still see some inconsistency with the use of these regularizations, and we should further investigate which conditions cause them to improve or degrade performance. We could also evaluate other data augmentation methods and which work better with Euclidean Alignment and with which each model and dataset. Finally, we could perform hyperparameter optimizations to improve the accuracy.

%In this work, we propose adding the Euclidean Alignment method directly to the EEG signals to increase the similarity of the domains, with the aim of improving the model's performance. A data augmentation process was also considered a form of regularization and expansion of the amount of data for training. The results allow us to observe a significant improvement up to 13.60\% with the addition of Euclidean Alignment in the shared models, which may be a good indication for a better generalization of the EEG signals. The data augmentation also proved effective in improving performance. The combination of these processes was effective and more powerful when used in conjunction with each other. For individual models, we still see some inconsistency with the use of these regularizations, with improvements and worsening depending on the architecture. Further studies are still needed, including combining EA with other data augmentation methods, switching the order of EA and S\&R, and tests on parameters such as $N_s$ values in a wider range of networks. Tests with larger datasets are also necessary to assess the architecture's generalization capacity.

\section{Acknowledgment}

GHR and RYC are grateful to São Paulo Research Foundation (FAPESP) for the financial support (grant 23/06715-3). The work of BA was supported by DATAIA Convergence Institute as part of the ``Programme d’Investissement d’Avenir'', (ANR-17-CONV-0003) operated by LISN.

\bibliographystyle{ieeetr}
\bibliography{references}  

\end{document}